\documentclass[sigconf]{acmart}
\renewcommand\footnotetextcopyrightpermission[1]{} 
\usepackage{enumerate}
\usepackage{booktabs}
\usepackage{multirow}
\usepackage[linesnumbered,ruled,vlined]{algorithm2e}
\usepackage{comment}
\usepackage{latexsym}
\usepackage{graphicx}
\usepackage{subfigure}
\usepackage{epstopdf}
\usepackage{epsfig}
\usepackage{xspace}
\usepackage[bottom]{footmisc}
\usepackage[10pt]{moresize}
\usepackage{todonotes}
\usepackage{url}
\SetKwRepeat{Do}{do}{while}
\newcommand{\mypara}[1]{
	\vspace*{0.01cm}
	\noindent\textbf{\textit{#1}}}
\newcommand{\RNum}[1]{\uppercase\expandafter{\romannumeral #1\relax}}

\begin{document}
\title{MiabNET: Message-in-a-bottle Protocol for MANET}

\author{Dongning Ma}
\email{dma2@villanova.edu}
\affiliation{%
  \institution{Department of Electrical and Computer Engineering, Villanova University}
}
	
\maketitle

\section{Introduction}
\label{sec:intro}
In this short paper, we propose MiabNET, a reactive protocol for Mobile Ad-hoc Networks (MANET). This protocol leverages the concept of ``message-in-a-bottle'' to spread the routing information though the entire network. The idea of the protocol is briefly described as below: if a node would like to find a route to a destination node not in the routing table, it will initialize a bottle and send this bottle to \textbf{a random one} of its neighbors. If this neighbor does not have the route to the destination, it will send the bottle to one of its random neighbors as well, until the bottle reaches the destination node.

While the bottle is drifting in the network, any node receiving this bottle will add itself to this bottle's travel history. This can prevent the bottle from stuck in a loop of several nodes and facilitate the route finding process. Moreover, by reading travel history in the bottle, the middle nodes in the network are able to know what nodes are accessible to them and correspondingly update their routing table. Therefore, one routing request from the node can help the entire network to be more optimized.

To prevent the bottle existing in the network forever, it has a designer-customizable limit of hops before being eliminated from the network. The source nodes also have a timer to calculate the time elapsed after sending out the bottle: time-outs will be recognized as that the bottle has been eliminated and the source nodes will send out a new bottle. A certain consecutive time-outs then indicates that the destination node is no longer accessible in the current network. 

In summary, the MiabNET protocol has the following features:
\begin{itemize}
    \item It is reactive: routing request will only be made when the source node cannot find the destination in its routing table.
    \item It allows network designer to control the use of bandwidth for route finding and can assign different routing priorities for different nodes.
    \item It is highly customizable and flexible for implementation of different flavor of it.
\end{itemize}

\section{Proposed Protocol Design}
\label{sec:proto}
\subsection{Assumptions}
We have the following assumptions for this protocol:
\begin{itemize}
    \item Connection between nodes are bi-directional.
    \item Each node has a unique and pre-determined ID.
    \item The node allows to have a limited buffer space to store routing information.
    \item The network allows routing packet size to vary.
\end{itemize}
\subsection{Attributes and Format}
To leverage MiabNET protocol, the nodes and packets should have the attributes (format) listed in Table.~\ref{tab:node} and Table.~\ref{tab:pkt} respectively.
\begin{table}[htbp]
  \centering
  \caption{Node attributes in MiabNET protocol.}
    \begin{tabular}{ccc}
    \toprule
    Name  & Type  & Description \\
    \midrule
    nid   & uint16 & Node ID \\
    bid   & uint16 & ID for the next gernerated bottle of this node \\
    nbors & list  & List of node IDs of neighbors \\
    rtab  & dict  & Routing table \\
    state & enum  & Current state of FSM of this node \\
    pkt\_queue & list  & List of packets to route \\
    btl\_queue & list  & List of bottles to manage \\
    \bottomrule
    \end{tabular}%
  \label{tab:node}%
\end{table}%

\begin{table}[htbp]
  \centering
  \caption{Packet format in MiabNET protocol}
    \begin{tabular}{ccc}
    \toprule
    Name  & Type  & Description \\
    \midrule
    src   & uint16 & Source node \\
    dest  & uint16 & Destination node \\
    btl\_id & str   & Unique identifier of this bottle \\
    rf    & bool  & Has this bottled found its destination? \\
    history & list  & List of nodes this bottle has been to \\
    \bottomrule
    \end{tabular}%
  \label{tab:pkt}%
\end{table}%

\subsection{Node Operation}
\begin{figure}
	\centering
	\includegraphics[keepaspectratio,width=0.5\columnwidth]{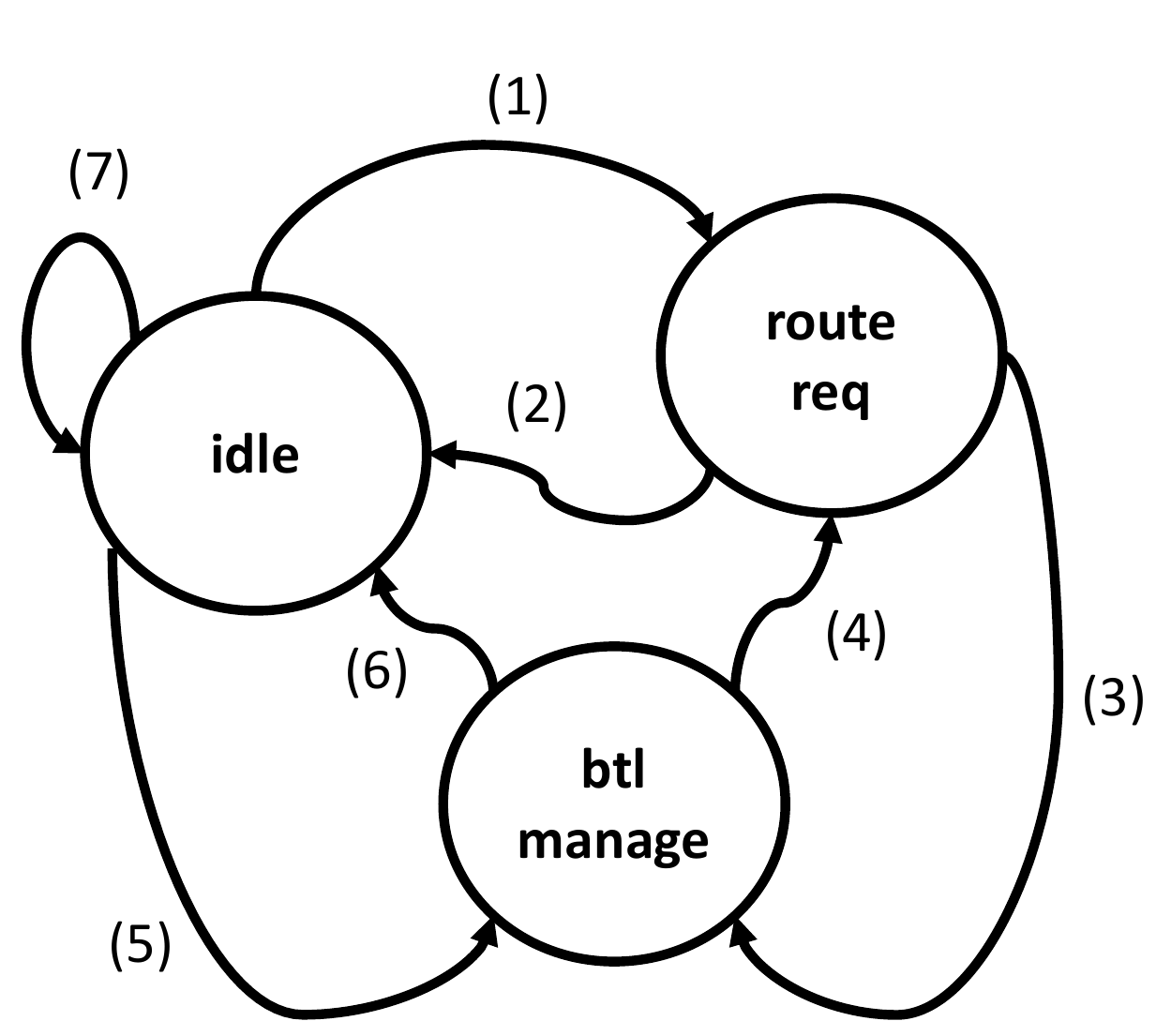}
	\caption{State transition graph for nodes with the MiabNET protocol.}
	\label{fig:fsm}
\end{figure}

The node is operated with a Mealy finite state machine (FSM). It has three states: idle, route\_req and btl\_manage. The transition of the states is described in Fig.~\ref{fig:fsm} where the transition conditions are listed below:
\begin{itemize}
    \item (1), (4): pkt\_queue is non-empty AND btl\_queue is empty 
    \item (2), (6), (7): pkt\_queue is empty AND btl\_queue is empty 
    \item (3), (5): btl\_queue is non-empty AND pkt\_queue is empty
\end{itemize}
\begin{figure*}
    \centering
    \subfigure[Network 1]{
        \includegraphics[keepaspectratio,width= 0.55\columnwidth]{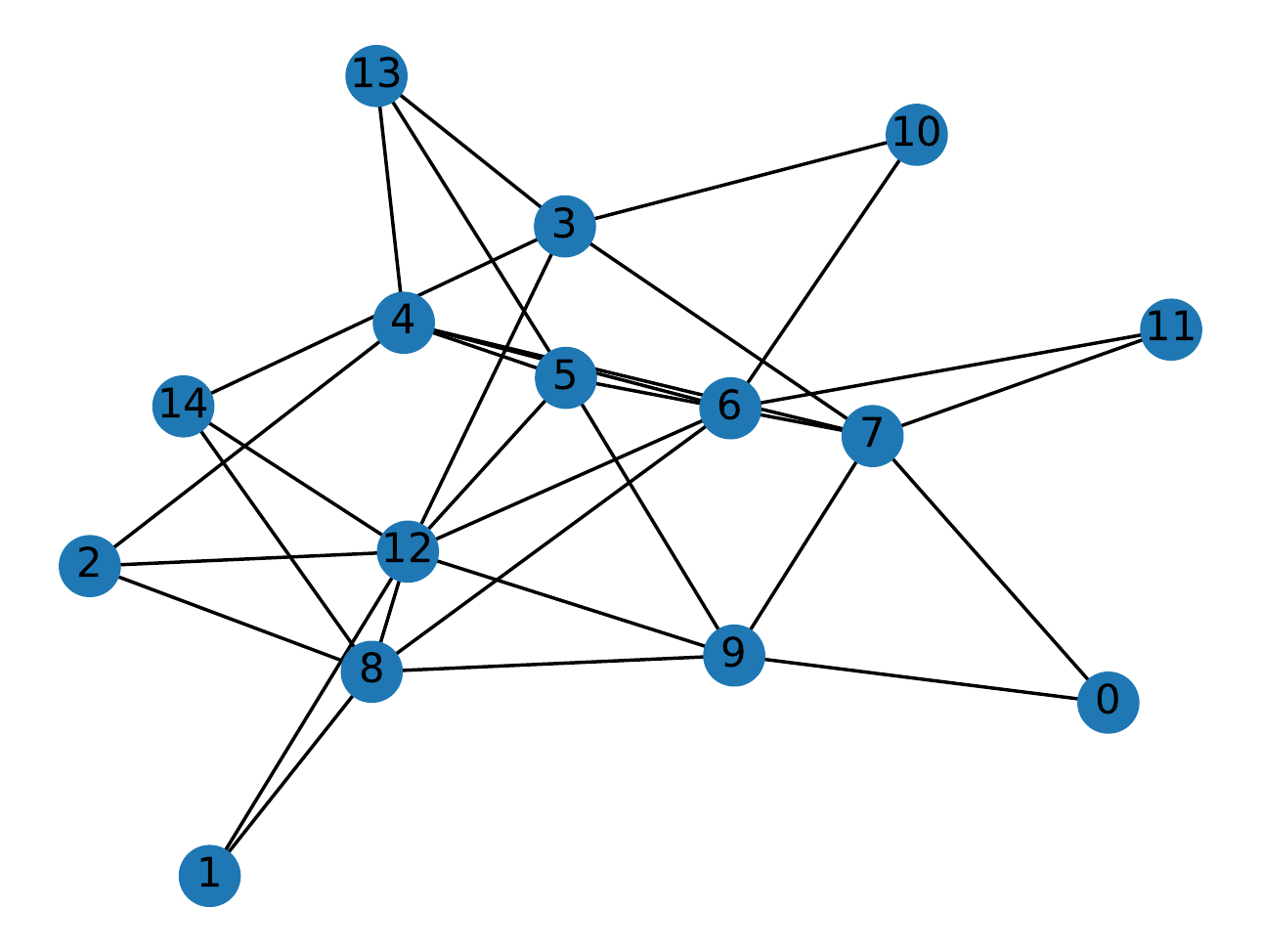}
    }
    \subfigure[Network 2]{
        \includegraphics[keepaspectratio,width= 0.55\columnwidth]{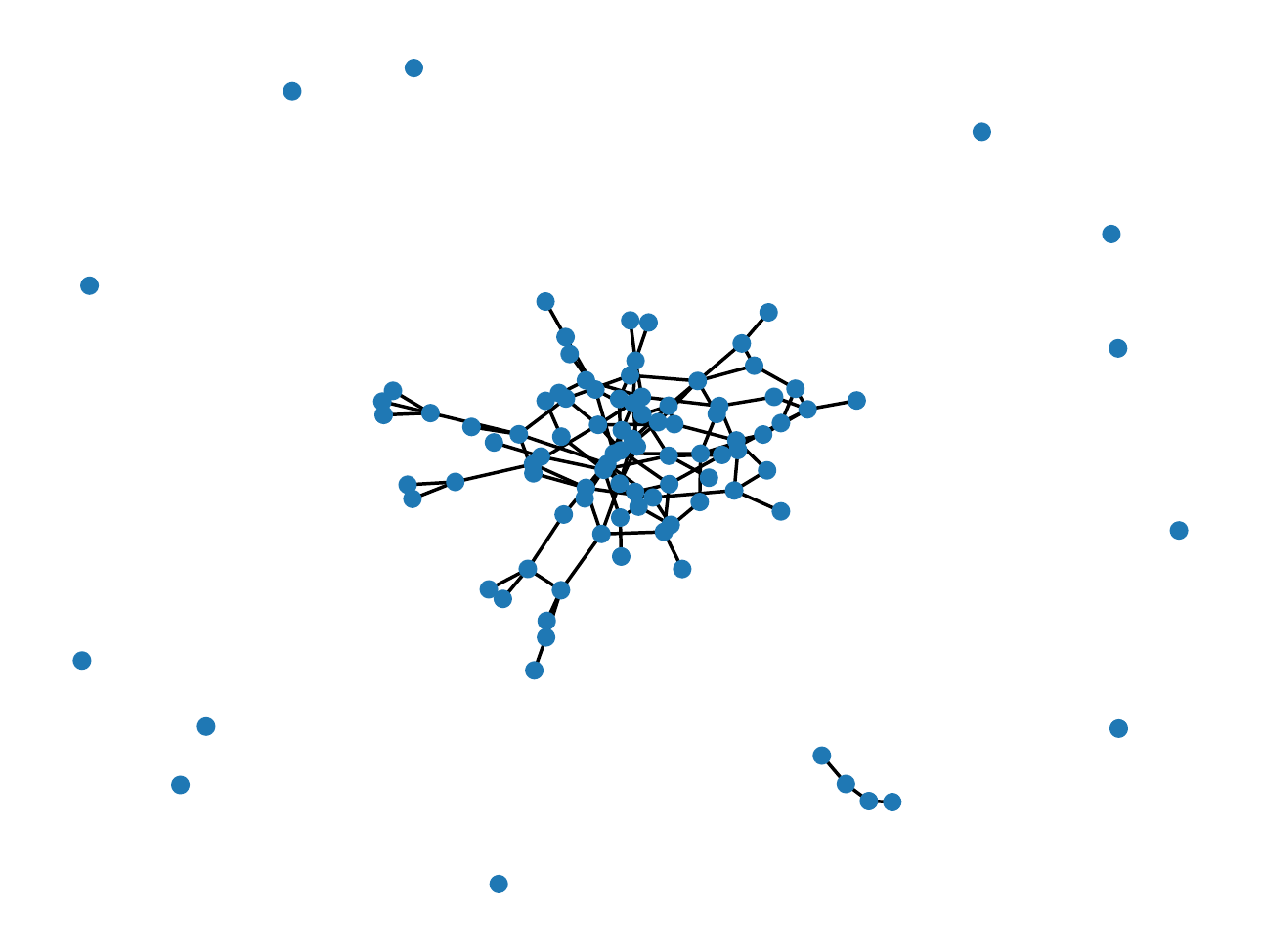}
    }
    \subfigure[Network 3]{
        \includegraphics[keepaspectratio,width= 0.55\columnwidth]{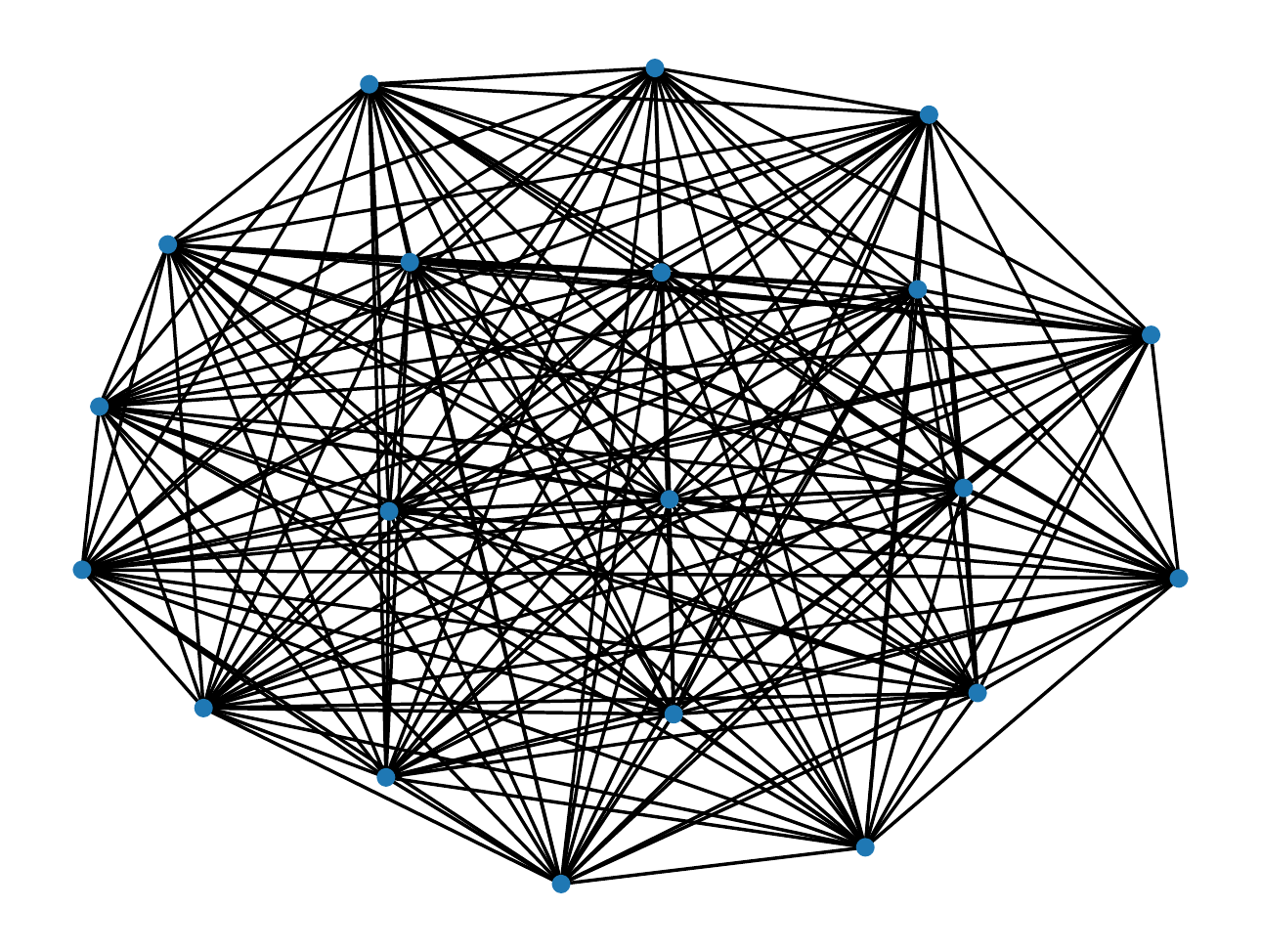}
    }
    \caption{Three example networks in simulation of MiabNET.}
    \label{fig:sim}
\end{figure*}
\mypara{idle} In the ``idle'' state, the node periodically checks if there are new routing requests (packet to send) or if there are bottles to manage by checking the two queues: pkt\_queue and btl\_queue. If either of these queue are non-empty, the node will enter the respective state of ``route\_req'' and ``btl\_manage''.

\mypara{route\_req}
Entering the ``route\_req'' state indicates that there are packet to send. If the destination node is already in the routing table, the packet can be immediately sent. However, if the destination is not in the routing table, the node will initialize a new bottle for routing request and send it to one of its random neighbors. The node will also set a timer and repeat limit. If the bottle does not come back before the time-out, the node can repeat the process above and initialize a new bottle. Additionally if the repeat times surpass the limit, the node will recognize the destination node as inaccessible. 

\mypara{btl\_manage}
Entering ``btl\_manage'' state indicates that the node received bottles from other nodes to manage. It is the most important state in this protocol as it defines how the knowledge in routing of the network is propagated. In this state there are three major tasks for a node. First is the routing table update. The node will read the travel history of the bottle and update the routing table if it finds a better route with fewer hops. If there are nodes that are not in the routing table, they will also be appended. Second is the bottle regulation. The node will check if the bottle has reached its maximum lifetime (hops limit). If so, the bottle will be eliminated. Third is the bottle forwarding. The node will check if it is the destination requested in the bottle, if so, it will set the boolean variable ``rf'' in the bottle to ``True'' to indicate ``route found''. When nodes receiving bottles marked as ``rf'', it will send back the bottle according to the travel history of the bottle. If the source node receives a ``rf''-marked bottle it previously sent, that means the bottle that successfully found the route has come back to the source node. The source node can thus update its routing table accordingly.

\subsection{Handling Exceptions}
The nodes obtain the information of their neighbors through periodic ``hello'' beacons. In this protocol, the failure of a packet delivery identifies a route or node failure and thus removes the entry of the routing table of the node trying to send packet to the failed nodes or through the failed route. The node that fails to send (delivery) the packet will send back the bottle with a special ``route failure'' message. In this case, the source node can initiate another route request by sending a new route request bottle. Therefore, the message complexity is O(N).

For node (re)connection, the routing table will not update unless the newly connected node initiates routing requests or other nodes initiate routing requests to the newly connected node. This is to minimize the overhead rought by this protocol.

\section{Simulation}
\label{sec:sim}
We implemented this protocol in python and simulate the proposed protocol with different scales of networks. In Fig.~\ref{fig:sim}, we shows the protocol applied in three networks. The first network is a small and generic network with 15 nodes. The second network is a huge, partitioned and sparse network, with 100 nodes. The third network is a strongly connected network with 20 nodes.

We randomly select two nodes as the source and destination nodes. Simulation results show that our proposed MiabNET protocol are able to find the route for all the three networks:

Network 1: [0, 7, 9, 12, 14, 3, 13, 4, 2, 8]

Network 2: [3, 93, 49, 60, 88, 57, 32, 76, 27, 12, 61, 33, 80, 39, 19]

Network 3: [16, 2, 1, 10, 18, 6, 7]

Though MiabNET is able to find route for all the three tested network, it has several drawbacks: First, for a completely newly constructed network, this protocol is in low efficiency, reflecting in higher delays, constant routing loops, etc. However, as more knowledge is propagated to more nodes, the routing efficiency will increase and the entire network will (slowly) converges toward optimization. Second, for larger networks, the overhead of the bottle is larger since it requires more space for storing its travel history. Third, the performance of MiabNET largely depends on the network structure. It suits better for sparse networks while weak at strongly connected networks.

\section{Conclusion}
In this paper, we propose MiabNET, a novel MANET protocol based on the idea of ``message-in-a-bottle''. This protocol is reactive and only requires limited overhead for route requesting. By helping source nodes find their destination, the middle nodes are also able to update their routing table, minimizing the hops required for forwarding packets to make the entire network more efficient. Simulation shows successful routing in three different types and scales of networks. Future work of this project will focus on simulation and performance evaluation on different network structures and also with different varying traffic patterns.

\end{document}